\documentclass[fleqn,usenatbib]{mnras}
\usepackage{newtxtext,newtxmath}
\usepackage[T1]{fontenc}
\usepackage{ae,aecompl}
\DeclareRobustCommand{\VAN}[3]{#2}
\let\VANthebibliography\thebibliography
\def\thebibliography{\DeclareRobustCommand{\VAN}[3]{##3}\VANthebibliography}
\usepackage{graphicx}
\usepackage{amsmath}
\usepackage{geometry}
\usepackage{ulem}
\usepackage{tabularx,booktabs}
\usepackage{caption}
\usepackage{dirtytalk}
\usepackage{multicol}
\usepackage{multirow}
\newcommand{\pcm}{\,cm$^{-2}$}	
\newcommand{\ps}{\,s$^{-1}$}	
\newcommand{\cps}{\,counts\,s$^{-1}$}	

\title[Burst analysis of GX 3$+$1 using \textit{AstroSat}]{Spectro-temporal and Type I X-ray burst analysis of GX 3$+$1 using \textit{AstroSat} observations}
\author[N. T. Thomas et al.]{
Neal Titus Thomas, S. B. Gudennavar\thanks{E-mail: shivappa.b.gudennavar@christuniversity.in} and S. G. Bubbly 
\\\\
Department of Physics and Electronics, CHRIST University, Bangalore-560029, India\\
}
\date{Accepted 2023 February 16. Received 2023 February 14; in original form 2022 July 25}
\pubyear{2023}
\begin{document}
\label{firstpage}
\pagerange{\pageref{firstpage}--\pageref{lastpage}}
\maketitle
\begin{abstract}
GX 3$+$1, an atoll type neutron star low-mass X-ray binary, was observed four times by Soft X-ray Telescope and The Large Area X-ray Proportional Counters on-board \textit{AstroSat} between October 5, 2017 and August 9, 2018. The hardness-intensity-diagram of the source showed it to be in the soft spectral state during all the four observations. The spectra of the source could be adequately fit with a model consisting of blackbody ($\mathtt{bbody}$) and power-law ($\mathtt{powerlaw}$) components. This yielded the blackbody radius and mass accretion rate to be $\sim$8 km and $\sim$2 $\times$ $10^{-9}$ M$_{\odot}$ y$^{-1}$, respectively. In one of the observations, a Type I X-ray burst having a rise and e-folding time of 0.6 and 5.6 s, respectively, was detected. Time-resolved spectral analysis of the burst showed that the source underwent a photospheric radius expansion. The radius of the emitting blackbody in GX 3$+$1 and its distance were estimated to be 9.19 $\substack{+0.97\\-0.82}$ km and 10.17 $\substack{+0.07\\-0.18}$ kpc, respectively. Temporal analysis of the burst yielded upper limits of the fractional RMS amplitude of 7$\%$, 5$\%$ and 6$\%$ during burst start, burst maximum and right after the radius expansion phase, respectively.
\end{abstract}
\begin{keywords}
X-rays: bursts --- stars: neutron --- X-rays: binaries --- accretion, accretion discs
\end{keywords}
\section{Introduction}
Low mass X-ray binaries hosting weakly magnetized neutron stars exhibit a wide range of spectral and temporal variability. Based on their correlated spectral and temporal variability properties, neutron star low mass X-ray binaries (NS-LMXBs) are classified into atoll and Z sources \citep{Hasinger1989}. The colour-colour-diagrams (CCDs)/hardness-intensity-diagrams (HIDs) of atoll sources are characterized by isolated clumps called ‘island states’ and elongated curved branches called the ‘banana branches' that are traced out on timescales of weeks to months. On the other hand, the CCDs/HIDs of Z sources are characterized by three main branches - horizontal branch (HB), normal branch (NB) and flaring branch (FB), which together form a Z shape that is traced out on timescales of hours to weeks \citep{Kuulkers1994}. Atoll sources have both lower neutron star magnetic-field strengths and lower accretion rates than Z sources, which can be linked to the differences in system parameters such as orbital period and properties of the donor star \citep{Hasinger1989}. Some atoll sources have been reported to exhibit Type I X-ray bursts, especially at luminosities below $2\times10^{37}$ erg~\ps \citep{Lewin1997}. These bursts are caused due to unstable hydrogen/helium burning in a thin shell on the surface of the neutron star \citep{Lewin1993}, and are characterized by a fast rise exponential decay (FRED) profile that lasts for a few seconds to tens of minutes.
\\[6pt]
GX 3$+$1 is a NS-LMXB discovered on June 16, 1964 by an \textit{Aerobee-Rocket} flight \citep{Bowyer1965}. It was identified to be an atoll type NS-LMXB based on the two branch structures (lower and upper banana branches) traced in its HID \citep{Lewin1987, Asai1993, Homan1998, Muno2002a}. There have been no conclusive reports on the island state of this source until now. GX 3$+$1 is one of the most luminous and persistently bright atoll sources associated with the bulge component of Milky Way Galaxy. Most bright X-ray sources in the galactic bulge exhibit soft thermal X-ray spectra (2.0 $-$ 10.0 keV), high X-ray luminosity ($10^{37} - 10^{38}$ erg~\ps), and moderate and irregular X-ray intensity variations. Some of these sources also exhibit thermonuclear X-ray bursts and rarely show kHz quasi-periodic oscillations (QPOs) \citep{Homan1998, Oosterbroek2001, Altamirano2010}.
\\[6pt]
Since its discovery, GX 3$+$1 has been observed by most of the major X-ray missions (\textit{EXOSAT} \citep{Schulz1989}, \textit{Ginga} \citep{Asai1993}, \textit{RXTE} \citep{Bradt1993, Kuulkers2000}, \textit{BeppoSAX} \citep{denHartog2003}, \textit{INTEGRAL} \citep{Paizis2006}, \textit{Chandra} \citep{vandenBerg2014}, \textit{XMM Newton} \citep{Pintore2015} and \textit{NuSTAR} \citep{Mondal2019}) to unravel its spectral and temporal properties. The optical counterpart of the source was identified as a 15.8 $\pm$ 0.1 mag star based on the studies of near-infrared spectrum of the companion \citep{vandenBerg2014}. The source distance was reported to be $\sim$6.1 kpc \citep{Kuulkers2000} from a double peak photospheric radius expansion (PRE) burst at a luminosity consistent with the Eddington luminosity limit for hydrogen-poor material. \citet{Galloway2008} determined a maximum source distance of 6.5 kpc to GX 3$+$1 assuming that Type I X-ray bursts are Eddington limited. During \textit{Hakucho} observations made between July - August, 1980, fifteen Type I X-ray bursts were observed from GX 3$+$1 \citep{Makishima1983}. The studies carried out with the observations using \textit{Granat} \citep{Molkov1999} and \textit{Ginga} \citep{Asai1993} observatories also revealed the Type I X-ray bursts from this source. Most of the Type I X-ray bursts observed from GX 3$+$1 are short with a decay time of 10 s. In addition to this, the source has also exhibited a long duration superburst having a decay time of 1.6 h \citep{Kuulkers2002}. Carbon left over from stable and unstable hydrogen and/or helium burning is thought to be a likely fuel for superbursts \citep{Cumming2001}. It has been reported that NS-LMXBs show the presence of large amplitude high coherence oscillations during various bursts \citep{Watts2012}, which are indicative of the spin frequency of the neutron star \citep{Markwardt1999,Strohmayer2008,Roy2021}. These oscillations have usually been detected between the rise and peak phase of the burst. However, no such oscillations have been detected from GX 3$+$1 so far. Another noteworthy temporal feature that NS-LMXBs exhibit in their power density spectra (PDS) is broadband noise which ranges from mHz to a few tens of Hz in frequency. Atoll sources in their 'island state' show the presence of a strong bandlimited noise component called the high frequency noise (HFN), which in most cases is flat at lower frequencies and steepens to an approximate power-law with an index of 1 at higher frequencies. However in the 'banana branch', the PDS of these sources are dominated by a power-law noise below 1 Hz, called the very low frequency noise (VLFN) \citep{Hasinger1989}.   
In 2003, \citet{denHartog2003} presented a comprehensive study of burst properties of GX 3$+$1 as a function of mass accretion rate using the long term observations of GX 3$+$1 from \textit{BeppoSAX} and ASM on \textit{RXTE}. The source was found to show an X-ray luminosity in the range $2\times10^{37}-4\times10^{37}$ erg \ps, with a maximum persistent bolometric luminosity of $6\times10^{37}$ erg \ps. 
\\[6pt]
\citet{Asai1993} reported that at timescales larger than one minute, GX 3$+$1 exhibits two kinds of variability. One has a time scale of the order of minutes to hours and the other one has a time scale of the order of years. At shorter timescales (minutes to hours), the HID of the source shows a correlation between the hardness and intensity, tracing out both upper and lower banana branches. However, on timescales of years, it was observed that the hardness was almost constant with variations only in the intensity. Further, the source exhibited strong long-term flux modulations of $\sim$29$\%$ of the average flux value on timescale of $\sim$6 yr \citep{Kotze2010}. 
\\[6pt]
Spectral studies on the source have shown that its X-ray spectrum can be reasonably well-fit by a model comprised of a black-body component, a Comptonised component and a Gaussian component \citep{Oosterbroek2001, Seifina2012}. \citet{Seifina2012} investigated the stability of spectral index as function of mass accretion rate using the observations from \textit{RXTE} and \textit{BeppoSAX}. It was demonstrated that the photon index was approximately constant ($\Gamma\approx2$) when the source moved from the faint phase to the bright phase, as well as during local transitions from lower banana branch to upper banana branch. The stability of the photon index was proposed to be an intrinsic signature of atoll source and was attributed to the dominance of the energy release in the Comptonization region with respect to the soft flux coming from the accretion disk. Further, GX 3$+$1 has shown the presence of a relativistically broadened Fe $K_{\alpha}$ line, which was first detected with \textit{BeppoSAX} \citep{Oosterbroek2001}. This is caused by the illumination of the accretion disk by the hard X-rays originating from the boundary layer/corona, which are reprocessed and re-emitted, causing a reflection spectrum. The reflection spectrum is modified by Doppler and relativistic effects which reveal properties of accretion flow in the system. The strength of these effects increases as the disk gets closer to the compact object, thus allowing the position of the inner accretion disk to be determined from the shape of the Fe $K_{\alpha}$ line proﬁle. Studies on the truncation of accretion disk at or prior to the surface of neutron star provide upper limits on its radial extent. The presence of a relativistically broadened Fe $K_{\alpha}$ line was confirmed by \textit{XMM-Newton} observations of the source. The profile of this feature suggested that the inner disk was located at a distance of $\sim$25 $R_{g}$, where, $R_{g}$ is the Schwarzschild or gravitational radius, with an inclination angle of $35^{\circ}-44^{\circ}$ \citep{Piraino2012}. Studies done by \citet{Pintore2015} using \textit{XMM-Newton} and \textit{INTEGRAL} observations suggested that the inner disk was closer to the neutron star at $\sim$10 $R_{g}$ with an inclination angle of $\sim35^{\circ}$, when accounting for the entire reﬂection spectrum.
\\[6pt]
Studies on NS-LMXBs are, in general, invaluable as they provide great insights into accretion geometry and the physical properties of the neutron star in the system. Moreover, analysis of the Type I X-ray burst from these systems helps to constrain several parameters of the system like radius of the neutron star in the system, Eddington flux, source distance etc. The literature on GX 3$+$1 reveals it to be an active burster with several X-ray bursts reported in the past. Erstwhile studies on the source have unveiled some physical properties of the system. But these studies had limited simultaneous spectral coverage in the soft and the hard bands. In addition, the previous distance estimate of the source (4.2$ - $6.4 kpc) reported by \citet{Kuulkers2000} has an uncertainty of 30$\%$, whereas, the one by \citet{Galloway2008} ($\sim$6.5 kpc) does not provide any uncertainties.
\\[6pt]
In order to explore the spectral and timing properties further and to put better constraints on its physical properties, GX 3$+$1 was studied with data obtained from the Soft X-ray Telescope (\textit{SXT}) and Large Area X-ray Proportional Counters (\textit{LAXPC}) on-board \textit{AstroSat}, observed over four observations between October 5, 2017 and August 9, 2018. The main focus of this paper is to investigate the spectral and temporal variations in the broadband emission from the source utilizing the simultaneous broadband spectral coverage capability of the \textit{SXT} and \textit{LAXPC}, and the unprecedented spectral and temporal resolution of \textit{LAXPC}. We present spectro-temporal analysis of the source carried out on the data from four observations. Additionally, we also report the results from time-resolved spectral analysis of a double peaked Type I thermonuclear X-ray burst seen in one of the observations, to study the evolution of various physical parameters of the source as a function of time. Using the results from spectral analysis, we infer various physical parameters i.e., radius of emitting blackbody, persistent flux, mass accretion rate, Eddington flux and source distance.
\\[6pt]
In the section that follows, data reduction and extraction procedure has been discussed. The procedure adopted for the broadband spectral, temporal and Type I X-ray burst analysis of GX 3+1 is presented in Section \ref{sec:spectral_temporal_X-ray burst_analysis}. Next, the results of this study and a comparison with the previously reported results on the source are discussed in Section \ref{sec:results_discussion}. Finally, inferences from the present study are summarised in Section \ref{sec:conclusions}. 
\section{Observations and Data reduction}
GX 3$+$1 was observed with the \textit{SXT} and \textit{LAXPC} for a total of four observations till date, on 05/10/2017 (hereafter, \textit{Observation 1}), 12/04/2018 (hereafter, \textit{Observation 2}), 29/04/2018 (hereafter, \textit{Observation 3}) and 09/08/2018 (hereafter, \textit{Observation 4}). The details of the observations are presented in Table \ref{Tab:Table1}. The \textit{SXT} instrument is a focusing telescope equipped with a charged coupled device camera. It has an effective area of 90 cm$^{2}$ at 1.5 keV and provides X-ray images in the 0.3 $-$ 8.0 keV energy range with a spectral resolution of $\sim$150 eV at 6 keV and a temporal resolution of $\sim$2.4 s in the photon counting (PC) mode and $\sim$278 ms in fast windowed (FW) mode \citep{Singh2016}. The \textit{LAXPC} instrument consists of three co-aligned proportional counters (\textit{LAXPC-10}, \textit{LAXPC-20}, \textit{LAXPC-30}). It operates in the 3.0 $-$ 80.0 keV energy range, with an effective area of $\sim$6000 cm$^2$, dead time of $\sim$42 $\mu$s and an absolute temporal resolution of 10 $\mu$s \citep{Yadav2016,Antia2017,Agrawal2017}. The PC and event analysis (EA) modes were employed for the observation of the source with \textit{SXT} and \textit{LAXPC}, respectively. 
\\[6pt]
Level 1 data of the source from \textit{SXT} instrument was processed using the standard \textit{SXT} pipeline - AS1SXTLevel2-1.4b\footnote{\url{https://www.tifr.res.in/~astrosat_sxt/sxtpipeline.html}} to obtain Level 2 event file for each orbit of an observation. These were then merged into one master event file using the \textit{SXT} Event Merger Tool\footnote{\label{note2}\url{https://www.tifr.res.in/~astrosat_sxt/dataanalysis.html}}. Further, the source image was extracted from the merged event file using XSELECT V2.4k. The source images for each observation were extracted from a circular region of $12.5^\prime$ radius centred at the source. The net count-rate in the energy range 0.3 $-$ 7.0 keV was found to be $\sim$38 \cps. Hence, the exclusion of the central source region to account for the pile-up was deemed not necessary. The background\footnote{SkyBkg\_comb\_EL3p5\_Cl\_Rd16p0\_v01.pha} and response matrix\footnote{sxt\_pc\_mat\_g0to12.rmf} files provided by the \textit{SXT} Payload Operations Centre (POC) were used for the analysis. Off-axis Auxiliary Response File (ARF) was created with the sxt\_ARFModule\footnote{sxtARFModulev03}. For all the observations, \textit{SXT} data in the range 0.7 $-$ 7.0 keV were used due to the poor data quality below 0.7 keV and above 7.0 keV. Level 1 data of the source from \textit{LAXPC} instrument was processed using LAXPCSOFT (Format A)\footnote{\url{http://astrosat-ssc.iucaa.in/laxpcData}} to obtain Level 2 event files, Good Time Interval (GTI) files, lightcurves, source and background spectra, Response Matrix Files (RMF) and PDS. Data from \textit{LAXPC-20} data alone was used to carry out spectral analysis as the performance of \textit{LAXPC-10} has not been good due to an abnormal change in its gain since 28$^{th}$ March, 2018 and \textit{LAXPC-30} was not operational during the \textit{Observations 2, 3} and \textit{4}. Temporal analysis was carried out with \textit{LAXPC-10}, \textit{20} and \textit{30} for \textit{Observation 1} and \textit{LAXPC-20} for \textit{Observations 2, 3} and \textit{4}. Further, as the data above 20.0 keV was dominated by background, the spectral and temporal studies with \textit{LAXPC} were restricted to 4.0 $-$ 20.0 keV energy range.
\section{Spectral, temporal and X-ray burst analysis}
\label{sec:spectral_temporal_X-ray burst_analysis}
\begin{figure}
    \centering
    \includegraphics[width=\columnwidth]{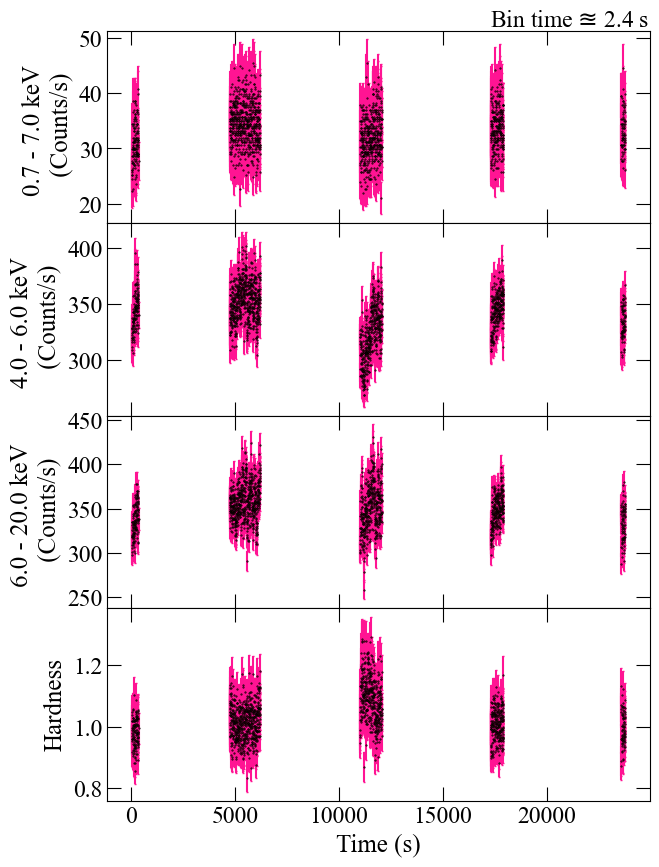}
    \caption{Simultaneous \textit{SXT} lightcurve in 0.7 $-$ 7.0 keV energy range (Panel 1 from top), \textit{LAXPC-20} net lightcurve in the energy ranges 4.0 $-$ 6.0 keV (Panel 2) and 6.0 $-$ 20.0 keV (Panel 3), and their hardness as function of time (Panel 4) from \textit{Observation 4}.}
    \label{fig:Lightcurve}
\end{figure}

\begin{figure}
    \centering
    \includegraphics[width=\columnwidth]{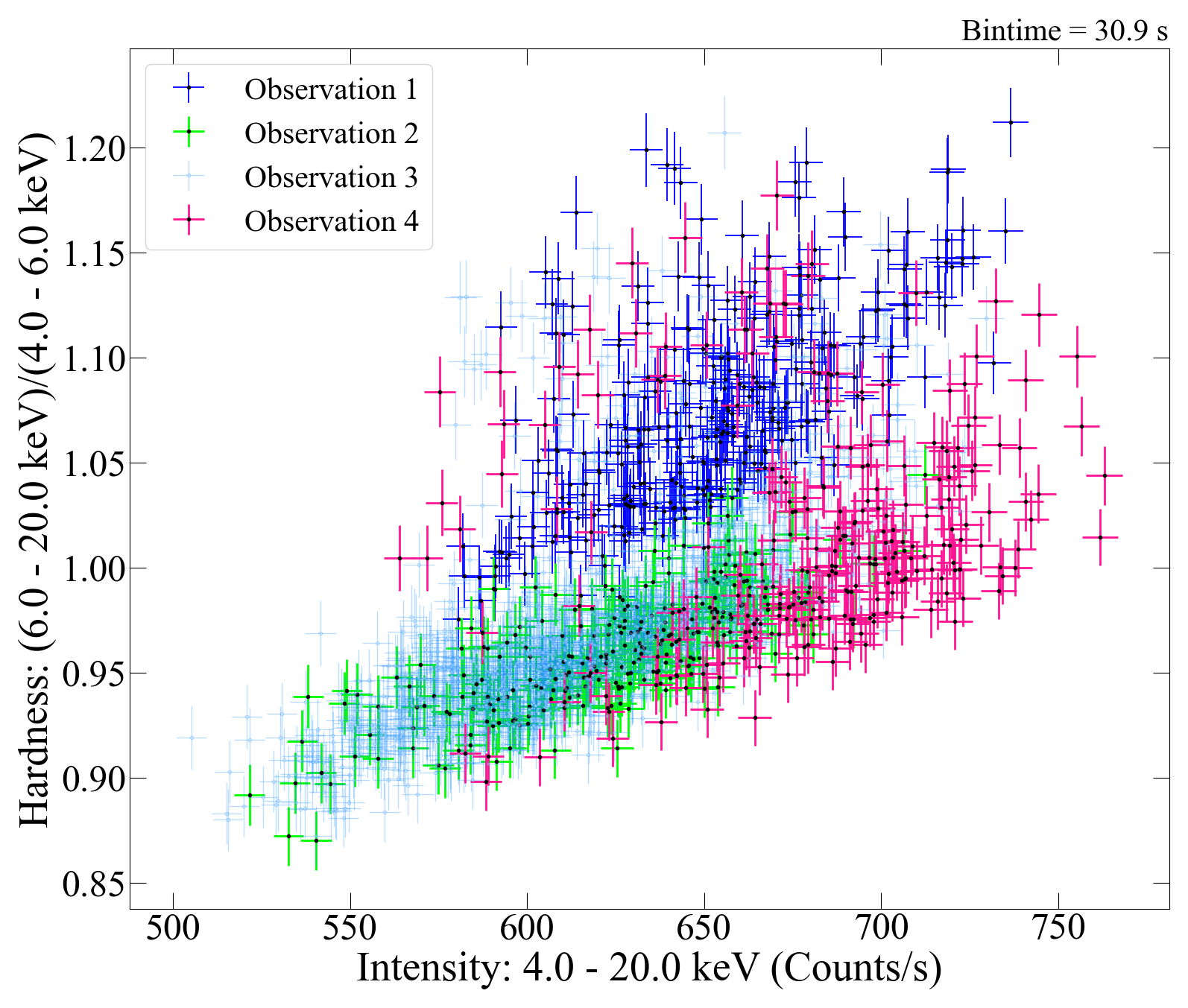}
    \caption{Merged HID of GX 3$+$1 with \textit{LAXPC-20} data.}
    \label{fig:HID}
\end{figure}

Net lightcurves of the source were obtained in the 0.7 $-$ 7.0 keV energy range from \textit{SXT}; and 4.0 $-$ 6.0 keV and 6.0 $-$ 20.0 keV energy ranges from \textit{LAXPC-20} with the bin size of $\sim$2.37 s. It was observed that the source intensity remained fairly constant at $\sim$35 \cps in the 0.7 $-$ 7.0 keV range; and $\sim$350 \cps in the 4.0 $-$ 6.0 keV and 6.0 $-$ 20.0 keV ranges during all the observations. The background subtracted lightcurves of \textit{Observation 4} from the \textit{SXT} and \textit{LAXPC-20} are shown in Figure \ref{fig:Lightcurve}. \textit{Observations 1, 2} and \textit{4} did not show presence of Type I X-ray bursts. However, a Type I X-ray burst was detected in \textit{Observation 3}. Burst analysis of this observation is presented in Section \ref{sec:burst_analysis}. \textit{LAXPC-20} data of the source was used to generate a combined $\sim$30 s HID of all the observations; with hardness defined as the ratio of counts \ps in 6.0 $-$ 20.0 keV to the counts \ps in 4.0 $-$ 6.0 keV energy range and intensity defined as the sum of counts \ps in 4.0 $-$ 20.0 keV energy range (Figure \ref{fig:HID}). The burst data was removed from \textit{Observation 3} before generating the HID. It is noticed that the hardness, which is related to its spectral shape, changes from $\sim$0.88 to $\sim$1.20; and the intensity varies from $\sim$500 to $\sim$750 \cps. A positive correlation between hardness and intensity is seen in the HID, which is characteristic of ‘banana branches' for atoll sources \citep{Hasinger1989, Asai1993}. From the fairly similar pattern traced in the HID by each observation of GX 3$+$1, it can be concluded that the source was in ‘banana branch' during all four observations. This is consistent with the results of \citet{Asai1993} and \citet{Mondal2019}. Further, the HID did not show any clue of ‘island states', which has not been exhibited by GX 3$+$1 until now.
\\[6pt]
In the following subsection, details of spectral analysis of all the observations are presented.
\begin{table*}
\centering
\caption{Observation log.}
\begin{tabular}{llllll}
\hline
Observation label & Observation ID & Date of observation & MJD & \multicolumn{2}{c}{Exposure time (ks)} \\
&                                  &  (dd-mm-yyyy)       &            &        \textit{SXT}    & \textit{LAXPC}\\
\hline
\textit{Observation 1} & G08\_039T01\_9000001582  & 05-10-2017 & 58031 & 2.91  & 9.57 \\
\textit{Observation 2} & G08\_039T01\_9000002036  & 12-04-2018 & 58220 & 7.58  & 8.05 \\ 
\textit{Observation 3} & A04\_122T01\_9000002064  & 29-04-2018 & 58237 & 28.21 & 41.15 \\   
\textit{Observation 4} & G08\_039T01\_9000002292  & 09-08-2018 & 58339 & 4.86  & 7.77 \\   
\hline
\end{tabular}
\label{Tab:Table1}
\end{table*}
\subsection{Spectral analysis}
\label{sec:spectral_analysis}
\begin{figure}
    \centering
    \includegraphics[width=\columnwidth]{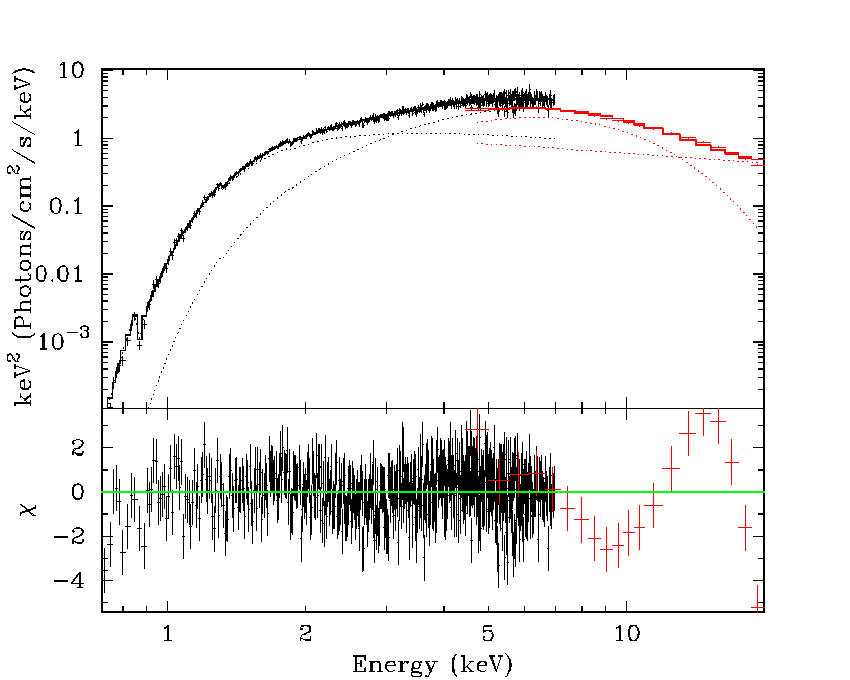}
    \caption{\textit{SXT} and \textit{LAXPC} combined spectrum of \textit{Observation 4} fit with the model combination $\mathtt{constant*tbabs(bbody+powerlaw)}$. The residuals ($\chi = (data-model)/error$) are plotted in the bottom panel.}
    \label{fig:Spectrum}
\end{figure}

Time averaged spectral analysis was carried out for each observation using the simultaneous broadband X-ray spectral coverage of \textit{AstroSat} with \textit{SXT} and \textit{LAXPC-20}. The simultaneous \textit{SXT} and \textit{LAXPC-20} spectra were obtained from the simultaneous GTI data of both the \textit{SXT} and \textit{LAXPC} instruments. Type I X-ray burst was removed from \textit{Observation 3}. The \textit{SXT} and \textit{LAXPC} spectra were jointly fit with the spectral modelling tool $\mathtt{XSPEC}$ $\mathtt{version: 12.10.1o}$ \citep{Arnaud1996}. The energy regimes below 0.7 keV and above 7.0 keV of \textit{SXT} were ignored due to uncertainties in its response and effective area. However, for \textit{LAXPC}, the energy regimes below 4.0 keV and above 20.0 keV were ignored due to presence of instrumental noise and large uncertainties in the background around  K X-ray fluorescence energy of Xe at 30 keV \citep{Antia2017}, respectively. The spectra of the source in the energy range 0.7 $-$ 20.0 keV could be adequately fit with a model combination containing a blackbody ($\mathtt{bbody}$) to fit the thermal emission; and a $\mathtt{powerlaw}$, to fit the non-thermal emission. The Tuebingen-Boulder Inter-Stellar Medium absorption model ($\mathtt{tbabs}$) with the solar abundance table given by \cite{Wilms2000} was used to account for absorption of source X-rays in the interstellar medium. In addition, a multiplicative constant factor was included in the model combination to address uncertainties caused due to cross calibration between the \textit{SXT} and \textit{LAXPC} instruments. As prescribed by the POC, a systematic error of 3$\%$ was added to all spectral fits \citep{Bhattacharya2017}. Further, a gain fit was performed with the slope of the gain frozen to unity and the offset left to vary to account for the non-linear change in the detector gain of the \textit{SXT} instrument. The model combination - $\mathtt{constant*tbabs(bbody+powerlaw)}$ yielded good fit for all the observations with reduced $\chi^2$ of $\sim$1.12. Positive residuals around 6.4 keV were not detected, indicating the absence of a reflection feature from the accretion disk. However, positive residuals were observed above 7.0 keV, which may be due to the underestimation of \textit{LAXPC} background. Further, $\mathtt{cflux}$ model was used to obtain the total unabsorbed flux in 0.7 $-$ 20.0 keV range. The best fit spectral parameters are given in Table \ref{Tab:Table2} and a sample best fit spectrum for \textit{Observation 4} is given in Figure \ref{fig:Spectrum}. It is to be noted that all values reported from spectral fits have a confidence level of 90$\%$. 
\begin{table*}
\centering
\caption{Best fit model parameters obtained from spectral analysis in the 0.7 $-$ 20.0 keV energy range.}
\begin{tabular}{llllll}
\hline
Model & Parameter & \textit{Observation 1} & \textit{Observation 2} & \textit{Observation 3} & \textit{Observation 4} \\ 
\hline
\\
$\mathtt{tbabs}$ & $N_H$ $(10^{22}$ cm$^{-2})$ & 3.08 $\substack{+0.11\\-0.10}$ & 3.01 $\substack{+0.06\\-0.09}$ & 2.96 $\substack{+0.08\\-0.06}$ & 3.01 $\substack{+0.07\\-0.10}$ \\\\
$\mathtt{bbody}$ & $kT$ (keV) & 1.57 $\substack{+0.03\\-0.02}$ & 1.48 $\pm$ 0.02 & 1.51 $\pm$ 0.02 & 1.57 $\pm$ 0.02 \\\\
& $norm$ $(10^{-2})$ & 5.18 $\substack{+0.12\\-0.13}$ & 5.11 $\pm$ 0.12 & 5.13 $\substack{+0.12\\-0.11}$ & 5.46 $\substack{+0.12\\-0.13}$ \\\\
$\mathtt{powerlaw}$ & $\Gamma$ & 2.59 $\pm$ 0.04 & 2.58 $\pm$ 0.03 & 2.57 $\pm$ 0.03 & 2.52 $\pm$ 0.03 \\\\
& $norm$ & 2.12 $\substack{+0.19\\-0.17}$ & 1.97 $\pm$ 0.12 & 1.9 $\substack{+0.13\\-0.09}$ & 2.01 $\substack{+0.13\\-0.15}$ \\\\
Unabsorbed total flux & ($10^{-8}$ erg cm$^{-2}$ s$^{-1}$) & 0.41 $\pm$ 0.01 & 0.4 $\pm$ 0.01 & 0.41 $\pm 0.01$ & 0.43 $\pm 0.01$\\\\
Reduced $\chi^2$ /dof & &1.00/529 & 1.04/610 & 1.22/641 & 1.23/572 \\\\
\hline
\end{tabular}
\label{Tab:Table2}
\end{table*}
From the normalization component ($norm$) of $\mathtt{bbody}$ model, source luminosity, $L$ (erg s$^{-1}$), was estimated using Eq. (1).
\begin{equation}
\begin{aligned}
norm = \frac{L_{39}}{(D_{10})^2}
\end{aligned}
\end{equation}
where, $D_{10}$ is the distance to the source (in units of 10 kpc) and $L_{39}$ is the source luminosity ($10^{39}$ erg s$^{-1}$). Radius of the blackbody, $R$ (km) was estimated using Eq. (2), 
\begin{equation}
\begin{aligned}
L = 4\pi R^{2} \sigma T^{4}
\end{aligned}
\end{equation}
where, $\sigma$ is the Stefan-Boltzmann constant (5.6704 $\times$ $10^{-5}$ erg cm$^{-2}$ s$^{-1}$ K$^{-4}$) and T is the absolute temperature (K) of the blackbody, that was determined using $kT$ (keV) value obtained from spectral analysis. The colour correction factor of $f_c = 1.48$ was obtained from \citet{Suleimanov2012} assuming that the source accreetes pure helium and has a luminosity of L $=$ 0.10$L_{Edd}$. Using this and assuming the source distance to be 6.1 kpc, obtained from the study of a PRE burst by \citet{Kuulkers2000}, the color corrected radius ($R_{*}$ $=$ $f_c R$) of the emitting blackbody was estimated to be in the range 7.14  $-$ 8.33 km. Further, mass accretion rate ($\dot{m}$) in units of g cm$^{-2}$ s$^{-1}$ and 10$^{-9}$ M$_{\odot}$ y$^{-1}$ at the surface of neutron star of GX 3$+$1 was estimated using the following relation (Eq. 3) \citep{Galloway2008},
\begin{equation}
\begin{aligned}
\dot{m} = \frac{6.7\times 10^3\times F_{p}^{\dagger}\times c_{bol} \left(1+z\right) D_{10}^2}{M_{NS}^{\dagger}\times R_{NS}^{\dagger}} 
\end{aligned}
\end{equation}
where, $F_{p}^{\dagger}$ is the total persistent flux in units of $10^{-9}$ erg cm$^{-2}$ s$^{-1}$ in the 0.7 $-$ 20.0 keV energy range, $c_{bol}$ is the bolometric correction for non-pulsating sources and its value is $\sim$1.38, $M_{NS}^{\dagger}$ and $R_{NS}^{\dagger}$ are the mass and radius of the neutron star in units of 1.4 M$_{\odot}$ and 10 km, respectively. $z$ is the surface redshift and $1+z =\left(1-\frac{2GM_{NS}}{R_{NS}c^2}\right)^{-1/2}$, G is the universal gravitational constant (cm$^{3}$ g$^{-1}$ s$^{-2}$) and c is the speed of light (cm s$^{-1}$). Assuming the neutron star mass to be 1.4 M$_{\odot}$ and its radius to be $\simeq R_{*}$, the mass accretion rate at the surface of the neutron star was calculated. It is to be noted that since the mass accretion rates derived throughout this work are calculated using band limited persistent flux,  these values are only lower limits. The values of radius ($R_{*}$), persistent flux ($F_p$) and mass accretion rate ($\dot {m}$) for all the observations are presented in Table \ref{Tab:Table3}.
\\[6pt] 
In the subsection below, details of temporal analysis of all the observations of the source are presented.
\begin{table*}
\caption{Radius, total persistent flux, mass accretion rate of GX 3$+$1 derived from the best fit parameters obtained from spectral analysis in the 0.7 $-$ 20.0 keV energy range.}
\begin{tabular}{lllll}
\hline
Observations & $R_{*}$ (km) & $F_{p}$ (10$^{-8}$ erg cm$^{-2}$ s$^{-1})$ & $\dot{m}$ ($10^4$ g cm$^{-2}$ s$^{-1}$) & $\dot {m}$ (10$^{-9}$ $M_{\odot}$ y$^{-1}$)\\
\hline
\\
\textit{Observation 1} & 7.33 $\substack{+0.10\\-0.19}$ & 0.41 $\pm$ 0.01 & 1.92 $\substack{+0.10\\-0.07}$ & 2.06 $\pm$ 0.02 \\\\
\textit{Observation 2} & 8.20 $\substack{+0.13\\-0.12}$ & 0.40 $\pm$ 0.01 &  1.68 $\pm$ 0.07 &  2.25 $\pm$ 0.02 \\\\
\textit{Observation 3} & 7.89 $\substack{+0.13\\-0.12}$ & 0.41 $\pm$ 0.01 &  1.79 $\pm$ 0.07 &  2.22 $\pm$ 0.02 \\\\
\textit{Observation 4} & 7.53 $\substack{+0.10\\-0.11}$ & 0.43 $\pm$ 0.01 &  1.96 $\pm$ 0.07 &  2.22 $\pm$ 0.02 \\\\
\hline
\end{tabular}
\label{Tab:Table3}
\end{table*}
\subsection{Temporal analysis}
\label{sec:temporal_analysis}
\begin{figure}
    \centering
    \includegraphics[width=\columnwidth]{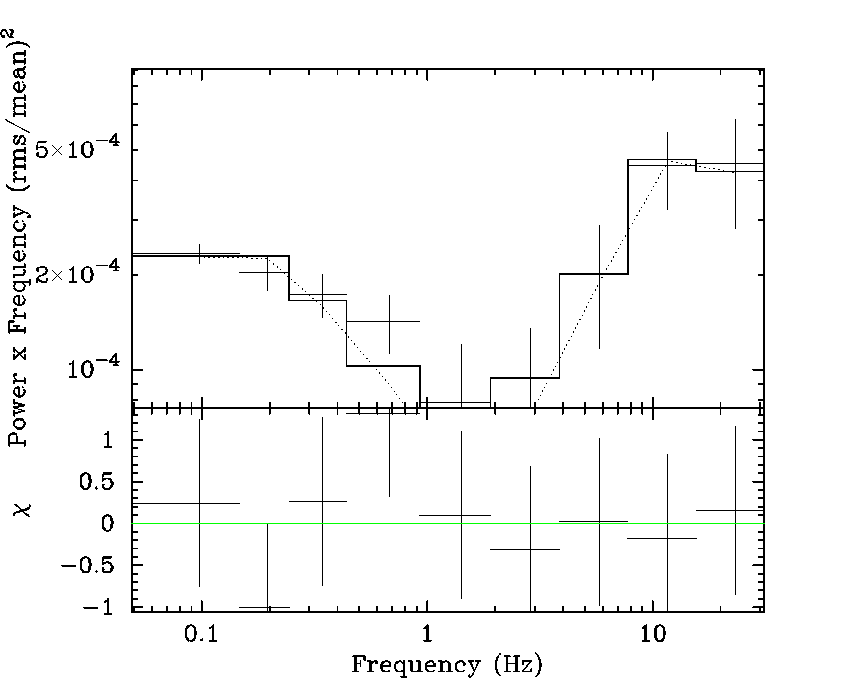}
    \caption{0.1 $-$ 50 Hz unfolded PDS from \textit{Observation 4} fit with Lorentzians. The residuals ($\chi = (data-model)/error$) are plotted in the bottom panel.}
    \label{fig:PDS}
\end{figure}

Temporal analysis was carried out in the 4.0 $-$ 20.0 keV energy range with data from  \textit{LAXPC-20} instrument in order to investigate the temporal properties of GX 3$+$1 during all the four observations. Root mean squared (RMS) normalized PDS in the 0.1 $-$ 100 Hz range was generated for each observation in the 4.0 $-$ 20.0 keV energy range. Temporal studies were restricted to 50 Hz throughout the analysis, as the power over 50 Hz was not significant. It was observed that the PDS for all observations were dominated by Poisson noise. Nevertheless, the PDS were modelled with a zero centered Lorentzian component to fit the broadband continuum feature with the addition of one or two more Lorentzian components as required to account for noise components. Narrow QPO features were not seen in the PDS and hence the PDS could be adequately fit with a combination of broad Lorentzians. The best-ﬁt parameters of temporal analysis for all four observations are presented in Table \ref{Tab:Table4}. A sample best fit PDS of \textit{Observation 4} is given in Figure \ref{fig:PDS}; where, the RMS normalized power is multiplied by frequency and plotted as a function of frequency \citep[see for e.g.,][]{Belloni1997,Nowak2000,Belloni2002}. This convention is used as it provides a better visual presentation of the distribution of power over Fourier frequencies. The results of temporal analysis show the presence of a broadband noise component which is present in all four observations of GX 3$+$1. 
\\[6pt]
In the next subsection, details of Type I thermonuclear X-ray burst analysis from \textit{Observation 3} are presented.
\begin{table*}
\centering
\caption{Best fit model parameters of PDS ((f)$-$ the parameter was frozen during the fit; *$-$ the error of that parameter could not be constrained).}
\begin{tabular}{llllll}
\hline
Model & Parameter & \textit{Observation 1} & \textit{Observation 2} & \textit{Observation 3} & \textit{Observation 4} \\ 
\hline
\\
$\mathtt{Lorentzian}$ $\mathtt{1}$ & $Line$ (Hz) & 0 (f) & 0 (f) & 0 (f) & 0 (f) \\\\
& $Width$ (Hz) & 0.18 $\substack{+0.03\\-0.04}$ & 0.19 $\pm$ 0.05 & 0.15 $\pm$ 0.03 & 0.23 $\substack{+0.06\\-0.05}$\\\\
& $norm$ $(10^{-3})$ & 0.77 $\pm$ 0.05 & 0.93 $\pm$ 0.09 & 0.76 $\substack{+0.06\\-0.05}$ & 0.82 $\pm$ 0.08\\\\
$\mathtt{Lorentzian}$ $\mathtt{2}$ & $Line$ (Hz) & 6.67 $\substack{+1.82\\-2.03}$ & 10.70 $\substack{+5.30\\-6.70}$ & 4.70 $\substack{+2.20\\-3.00}$ & 11.00 $\substack{+5.30\\-6.20}$ \\\\
& $Width$ (Hz) & $20.0^*$ & $20.0^*$ & $20.0^*$ & $20.0^*$ \\\\
& $norm$ $(10^{-3})$ & 0.84 $\substack{+0.10\\-0.09}$ & 1.10 $\pm$ 0.40 & 1.20 $\substack{+0.20\\-0.10}$ & 1.10 $\pm$ 0.30 \\\\
$\mathtt{Lorentzian}$ $\mathtt{3}$ & $Line$ (Hz) & 0.57 $\substack{+0.10\\-0.17}$ & - - - & - - - & - - - \\\\
& $Width$ (Hz) & 0.3 $\substack{+0.4\\-0.2}$ & - - - & - - - & - - - \\\\
& $norm$ $(10^{-3})$ & 0.05 $\substack{+0.02\\-0.03}$ & - - - & - - - & - - - \\\\
Reduced $\chi^2$ /dof & &1.18/10 & 1.19/5 & 1.53/8 & 0.78/4 \\\\
\hline
\end{tabular}
\label{Tab:Table4}
\end{table*}
\subsection{Spectral and temporal analysis of Type I thermonuclear X-ray burst}
\label{sec:burst_analysis}
\begin{figure}
    \centering
    \includegraphics[width=\columnwidth]{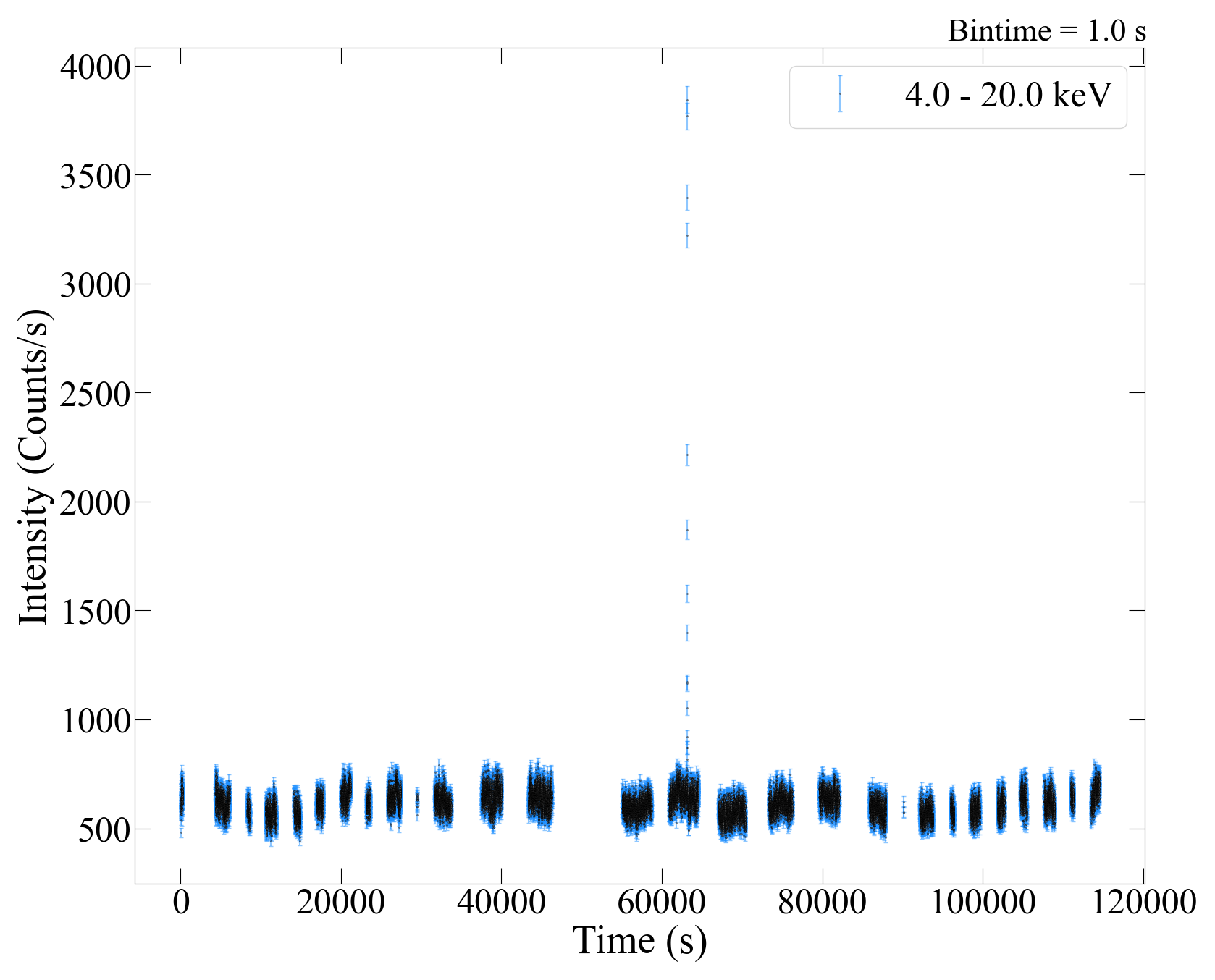}
    \caption{1 s binned \textit{LAXPC-20} net lightcurve in the 4.0 - 20.0 keV energy range of Type I X-ray burst seen in \textit{Observation 3}.}
    \label{fig:burst_LAXPC_lightcurve}
\end{figure}

\begin{figure}
    \centering
    \includegraphics[width=\columnwidth]{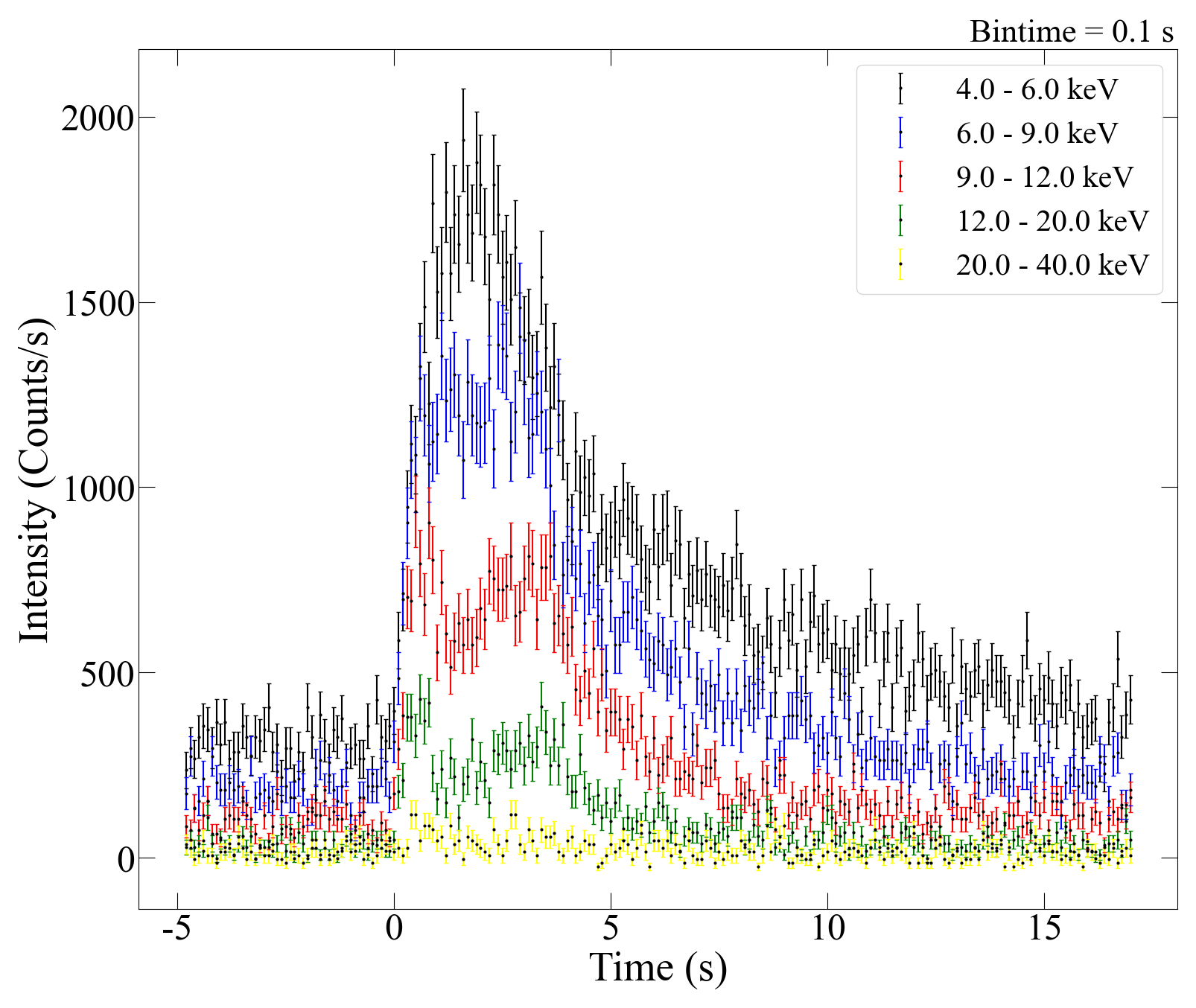}
    \caption{0.1 s binned energy resolved net lightcurves of Type I X-ray burst seen in \textit{Observation 3}.}
    \label{fig:burst_lightcurve_profile}
\end{figure}

\begin{figure}
    \centering
    \includegraphics[width=\columnwidth]{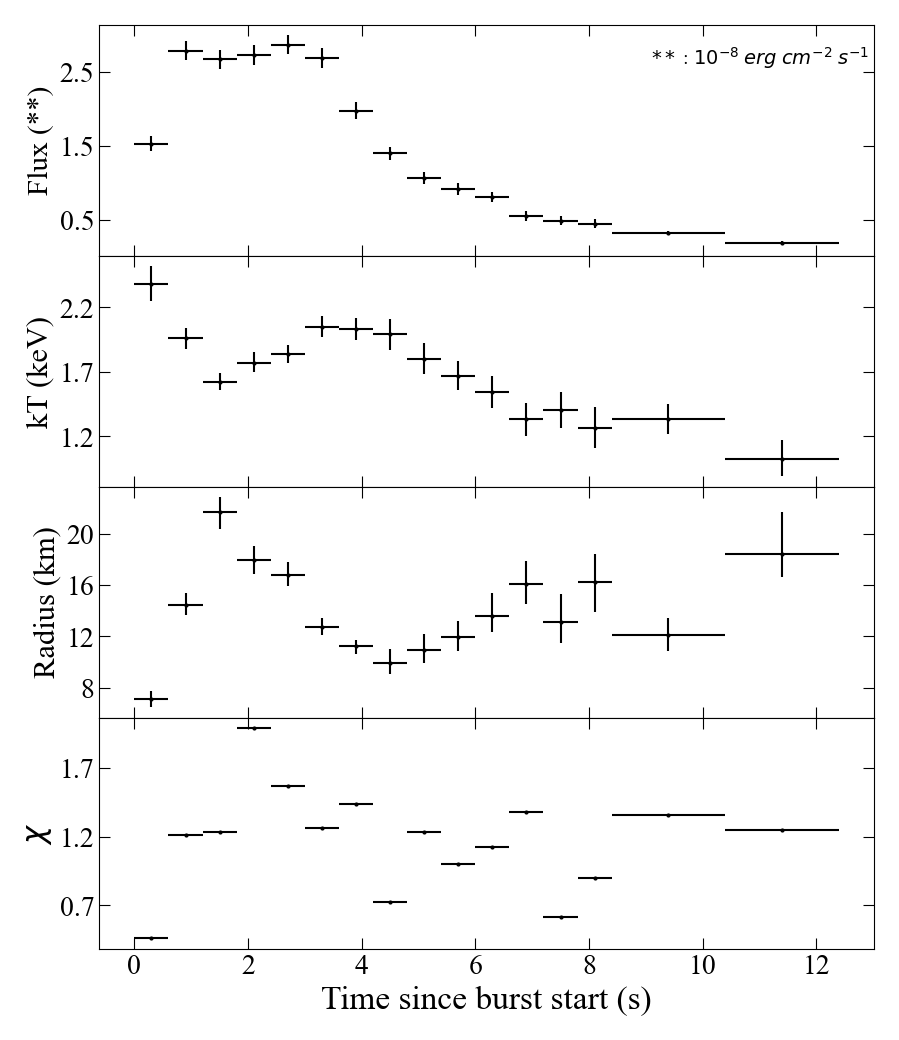}
    \caption{Variation of best ﬁt spectral parameters during Type I thermonuclear X-ray burst from GX 3$+$1 detected in \textit{Observation 3}. Evolution of the burst flux (Panel 1, from top), blackbody temperature ($kT$) (Panel 2), colour corrected blackbody radius ($R_{*}$) and (Panel 3) and residuals ($\chi = (data-model)/error$) (Panel 4).}
    \label{fig:Time_resolved_spectral_study}
\end{figure}

A Type I thermonuclear X-ray burst from GX 3$+$1 was detected in \textit{Observation 3}. The X-ray burst was outside the good time intervals of the \textit{SXT} instrument and was detected only in the \textit{LAXPC} data (Figure \ref{fig:burst_LAXPC_lightcurve}). The energy resolved burst profile, obtained from \textit{LAXPC-20} data, of the burst in the energy ranges: 4.0 $-$ 6.0 keV,  6.0 $-$ 9.0 keV, 9.0 $-$ 13.0  keV, 13.0 $-$ 20.0 keV and 20.0 $-$ 40.0 keV, shows that the burst is significant up to 20.0 keV (Figure \ref{fig:burst_lightcurve_profile}). The burst lightcurve in the 4.0 $-$ 6.0 keV range shows a canonical FRED profile. However, in energy 6.0 $-$ 9.0  keV, 9.0 $-$ 13.0 keV and 13.0 $-$ 20.0 keV ranges, the burst exhibits a double peak shape. The double peak shape of Type I thermonuclear X-ray burst has been shown to be linked with PRE \citep{Lewin1984}. The characteristics of the burst were computed using a 0.1 s net lightcurve in the 4.0 $-$ 20.0 keV energy range following methods similar to the one used by \citet{Galloway2008}. The start time of the burst was defined as the time-bin prior to the time when burst count-rate first exceeded 25$\%$ of the peak count-rate. The rise time, defined as the time interval from the start time to when the burst count-rate exceeded 90$\%$ of the peak count-rate, was computed to be 0.6 s. Further, the e-folding time of the burst was found to be 5.6 s. The spectral parameters of NS-LMXBs evolve with time during rapid events like bursts. Therefore it is important to divide burst into several short time intervals and fit the spectrum from each interval, from the rise to the decay of the burst, to understand the time evolution of the spectral parameters \citep{Lewin1993}. Time-resolved spectral analysis was carried out in order to study the variation of spectral parameters of the burst as a function of time. A pre-burst spectrum of the source was extracted following a method similar to the one used by \citet{Gruver2022}. Pre-burst spectrum from only the top layer (to improve counting statistics) of \textit{LAXPC-20}, having 100 s exposure time, 200 s prior to the burst, was extracted and fit with the model combination $\mathtt{tbabs(bbody+powerlaw)}$ available in the spectral modeling tool $\mathtt{XSPEC}$ $\mathtt{version: 12.10.1o}$ \citep{Arnaud1996}. The value of $\mathtt{N_H}$ parameter in $\mathtt{tbabs}$ model was frozen to 2.96 $\times$ $10^{22}$ \pcm, obtained from the spectral modeling of the continuum of \textit{Observation 3} (Table \ref{Tab:Table2}). Energy range for the spectral modeling was restricted to 4.0 $-$ 20.0 keV due to high residuals below 4.0 keV and high background above 20.0 keV. The spectral fit yielded a reduced $\chi^2$ /dof = 1.22/16 with a $\mathtt{bbody}$ temperature, $kT$ (keV) and $norm$ of 1.67 $\substack{+0.09\\-0.07}$ and 3.74 $\substack{+0.49\\-0.42}$ $\times 10^{-2}$, respectively; and $\mathtt{powerlaw}$ photon index ($\Gamma$) and \textit{norm} of 2.91 $\substack{+0.17\\-0.19}$ and 5.93 $\substack{+2.98\\-2.27}$, respectively. This yielded a pre-burst flux of 0.23 $\pm 0.03$ $\times$ 10$^{-8}$ erg cm$^{-2}$ s$^{-1}$ in the 4.0 $-$ 20.0 keV energy range. Further, X-ray spectra from the top layer of \textit{LAXPC-20}, having exposure time of 0.6 s until the burst e-folding time and exposure time of 2 s from then on, till the rest of the burst duration, were extracted and fit using two methods i.e., classical method and the variable persistent flux method \citep{Worpel2013,Worpel2015}. The classical method involves fitting the burst spectra with a fixed pre-burst spectral model along with the addition of blackbody model ($\mathtt{bbody}$). On the other hand, the variable persistent flux method uses a scaling factor $f_a$ multiplied to the fixed pre-burst model in addition to a blackbody model to get statistically better fits. The factor $f_a$ accounts for the variation of the persistent emission during the burst. However, since this approach could not yield errors on the blackbody temperature and \textit{norm} in certain segments of the burst, the results obtained with the classical method were adopted. Figure \ref{fig:Time_resolved_spectral_study} shows the evolution of the temperature, flux and radius of the blackbody corrected with the colour correction factor $f_c$ $=$ 1.48 \citep{Suleimanov2012}  obtained from time-resolved spectral analysis with the classical approach. Examination of the evolution of the best-fit spectral parameters during the burst shows a rapid increase in the radius of the emitting blackbody accompanied by a drop in the blackbody temperature. This confirms that the source has undergone a PRE during the burst. The photosphere touches down to the surface of the neutron star at 4.5 s after the start time of the burst. The Eddington luminosity (erg s$^{-1}$) of the source measured by an observer at infinity was calculated using Eq. (4) \citep{Galloway2008}.
\begin{equation}
\begin{aligned}
L_{Edd,\infty} = 2.7 \times 10^{38} \times M_{NS} \times \frac{1 + (\alpha_{T} T)^{0.86}}{(1 + X)(1+z)}
\end{aligned}
\end{equation}
where, $\alpha_T$ is a coefficient that parameterizes the temperature dependence of the electron scattering opacity ($\simeq$ 2.2 $\times$ $10^{-9}$ $K^{-1}$ \citep{Lewin1993}), T is the effective temperature (K) of the atmosphere and X is the mass fraction of hydrogen in the atmosphere. Assuming the burst to radiate isotropically and the burst fuel to be pure helium (X = 0), the Eddington luminosity at infinity was found to be 2.7 $\substack{+0.02\\-0.09}$ $\times$ 10$^{38}$ erg s$^{-1}$. Following this, the distance to the source was calculated to be 8.84 $\substack{+0.04\\-0.21}$ kpc using Eq. (5).
\begin{equation}
\begin{aligned}
d =  \biggl(\frac{L_{Edd,\infty}}{4 \pi \xi_b F_{pk}}\biggl)^{1/2}
\end{aligned}
\end{equation} 
where, $\xi_b$ is the anisotropy constant for the burst flux (as given in Eq. 6), assumed to be 1 for isotropic emission, $F_{pk}$ = 2.87 $\pm$ 0.13 $\times$ $10^{-8}$ erg cm$^{-2}$ s$^{-1}$ is the observed peak flux of the PRE burst in the 4.0 $-$ 20.0 keV energy range. However, it is likely that the angular distribution of the burst emission is not uniform. Hence, it becomes necessary to introduce anisotropy in the calculation of source distance. The anisotropy constant for the burst flux, $\xi_b$, is given by \citep{He2016}, 
\begin{equation}
\begin{aligned}
\xi_b^{-1} = \xi_d^{-1} + \xi_r^{-1}
\end{aligned}
\end{equation} 
where $\xi_d$ and $\xi_r$ are the anisotropy constant for direct burst flux and reflected burst flux, respectively, and given as in Eqs. (7) and (8), respectively.
\begin{equation}
\begin{aligned}
\xi_d^{-1} =  \frac{1 + |\cos \theta|}{2}
\end{aligned}
\end{equation} 
\begin{equation}
\begin{aligned}
\xi_r^{-1} =  \frac{|\cos \theta|}{2}
\end{aligned}
\end{equation} 
where, $\theta$ is the inclination angle of the source disk. Considering a disk inclination of $35^{\circ}$\citep{Pintore2015} $\xi_b$ was calculated to be 0.76 and was used in Eq. (5) to obtain a source distance of 10.17 $\substack{+0.07\\-0.18}$ kpc.
\\[6pt]
To the best of our knowledge, burst oscillations have not been reported from GX 3$+$1 until now. However, such oscillations have been reported in several other atoll sources \citep{Strohmayer2008,Galloway2008,Galloway2020}. In order to investigate if GX 3$+$1 exhibits such a temporal signature, an in-depth temporal analysis was performed. This was done adopting the Fast Fourier Transform (FFT) technique to generate 10 $-$ 1000 Hz, Leahy normalized power spectra in the 4.0 $-$ 20.0 keV energy range for 1 s window moving forward in steps of 0.1 s for 6 seconds from the start of the burst. Examination of the Leahy normalized power spectra, showed local maxima at $\sim$317 Hz during rise of the burst and $\sim$338 Hz after the touchdown phase, which may indicate coherent burst oscillations. The significance of these detections were computed using the relation $x = e^{-P_{\max}/2} \times n$, where $P_{\max}$ is the maximum Leahy power measured and $n$ is the number of trials over which the detection was made. The confidence level of the detection is given by X$\sigma$, where X = $\sqrt{2}$ erf$^{-1}(1 - x)$ \cite[see e.g.][]{Roy2021}. The details of the burst oscillation candidates are given in Table \ref{Tab:Table5}. Considering 50 trials, the confidence levels of the detection of both burst oscillation candidates were calculated to be $\sim$2.5$\sigma$. This low significance may be due to poor signal to noise ratio, owing to availability of only \textit{LAXPC-20} data. However, it could also mean that these oscillations are merely noise peaks. Following \citet{Kuulkers2000}, the upper limit on the fractional RMS amplitude of the oscillations was also calculated. The fractional RMS amplitude ($A$) of the oscillation is given by $A$ $=$ $\sqrt{P_{UL}/{N}}$, where $P_{UL} = P_{\max} - P_{exceed}$ is the upper limit to the signal and N is the number of photons in the interval during which the signal was measured \citep[see Section 4.2.2 of][]{vanderKlis1988}. Using 1s long burst segments in the 4 $-$ 20 keV range, the upper limit of the fractional RMS amplitude during the burst start, burst maximum and right after the radius expansion phase, was calculated to be 7$\%$, 5$\%$ and 6$\%$, respectively. In Section \ref{sec:results_discussion}, the results obtained from our study are discussed and compared with the previous studies on GX 3$+$1. 
\begin{figure*}
    \centering
    \includegraphics[width=\columnwidth]{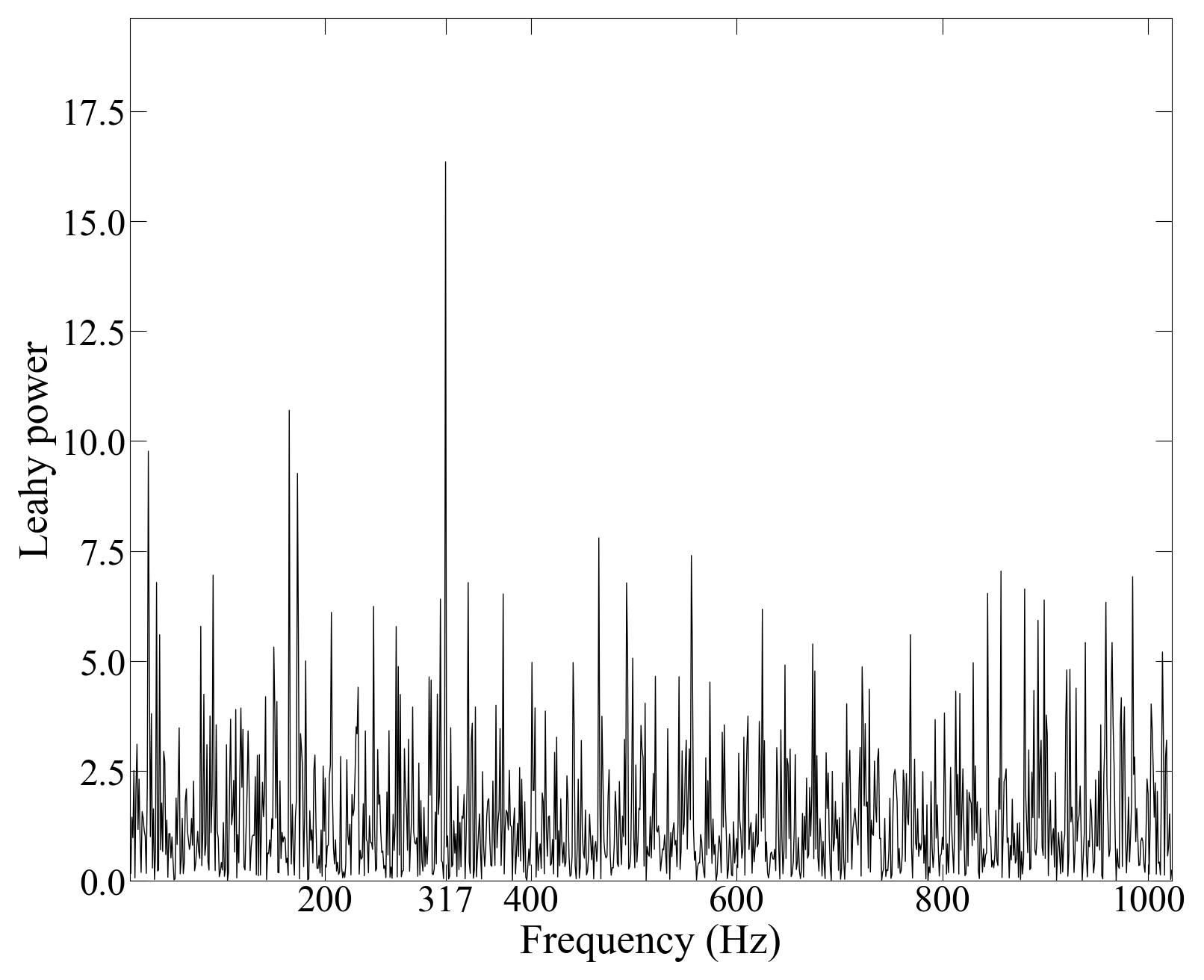}
    \includegraphics[width=\columnwidth]{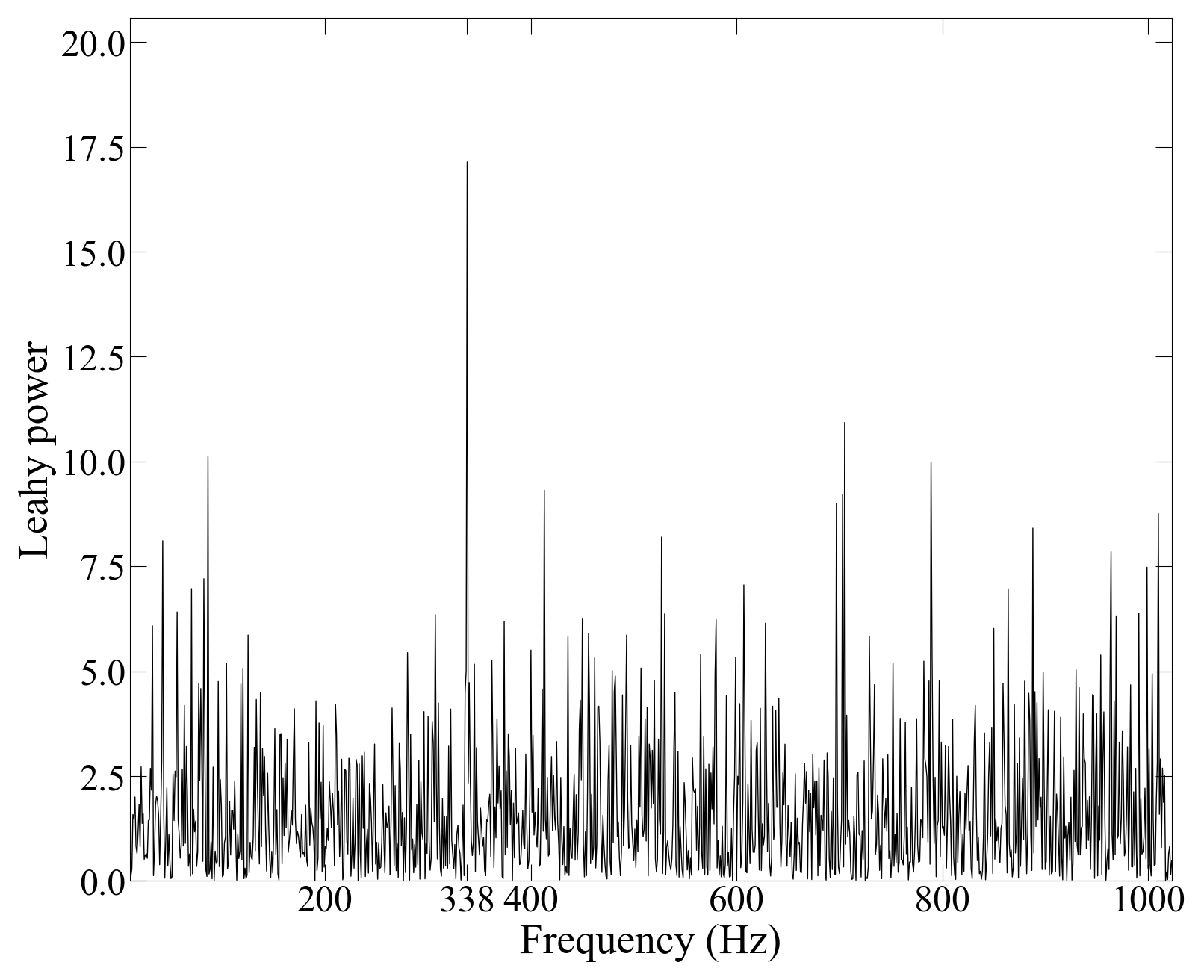}
    \caption{The Leahy power spectrum, for a 1s window, in 4 $-$ 20 keV range, of the burst oscillation candidate at 317 Hz (left panel) and 338 Hz (right panel).}
    \label{Fig:burst_oscillation_candidates}
\end{figure*}
\begin{table}
\centering
\caption{Detection phase, maximum measured Leahy power and confidence level of the burst oscillation candidates.}
\begin{tabular}{llll}
\hline
Detection phase & Oscillation frequency (Hz) & $P_{\max}$ & Confidence level \\
\hline
Rise        & 317 & 16.34  & 2.5$\sigma$ \\
Touchdown  & 338 & 17.14  & 2.6$\sigma$ \\
\hline
\end{tabular}
\label{Tab:Table5}
\end{table}
\section{Results and Discussion}
\label{sec:results_discussion}
In this work, we report the broadband spectral and temporal analysis of the atoll NS-LMXB GX 3$+$1 in the energy ranges 0.7 $-$ 20.0 keV and 4.0 $-$ 20.0 keV, respectively, using data from \textit{SXT} and \textit{LAXPC-20} over four observations. HID of the source (Figure \ref{fig:HID}) shows a positive correlation between hardness and intensity indicating that the source was on the banana branch of the atoll track during the four observations. Considering that the observation epochs used in this study are of short timescale, this behaviour agrees with the one reported by \citet{Asai1993}, where the spectral colours vary such as to move between the lower and upper banana branches on timescales of a few days. The presence of the source in banana branch is also consistent with previous studies  \citep{Oosterbroek2001,Mondal2019}, which have reported GX 3$+$1 to only trace the banana branch in its HID. 
\\[6pt]
In order to understand the spectral properties of GX 3$+$1, its broadband spectra in the 0.7 $-$ 20.0 keV energy band using data from \textit{SXT} and \textit{LAXPC}, were investigated. The \textit{SXT} and \textit{LAXPC} joint spectra of all four observations could be adequately modelled with a simple model combination containing an absorbed $\mathtt{blackbody}$ along with a $\mathtt{powerlaw}$ (Figure \ref{fig:Spectrum}). The origin of the blackbody component has been thought to be from either the neutron star surface or the accretion disc \citep{Gilfanov2003}. Positive residuals are seen at energies greater than 10 keV. This is seen in several accreting NS-LMXBs as reported by \citet{DiSalvo2000,DiSalvo2001,DiSalvo2006,Paizis2006,Piriano2007}. It has also been reported in a previous study on this source conducted by \citet{Pintore2015}. These residuals may correspond to the Compton back-scattering hump as reported by \citet{Mondal2019}. However, the reflection feature at $\sim$6.4 keV reported by them is not found in the present study. The blackbody temperature (\textit{kT}) obtained from the spectral fits is observed to vary from 1.48 to 1.58 keV. Moreover, the consistency of the spectral parameters from all four observations is in agreement with the similar pattern traced by these observations in the HID as well. Further, the colour corrected radius ($R_*$) of the emitting blackbody calculated from the normalization value (\textit{norm}) of the blackbody model ($\mathtt{bbody}$) assuming a distance of 6.1 kpc \citep{Kuulkers2000}, as done by \citet{Mondal2019}, is in the range 7.93 $-$ 8.86 km (Table \ref{Tab:Table3}). In addition to this, a better estimate of the blackbody radius is obtained from time-resolved spectral analysis of the Type I thermonuclear X-ray burst exhibited by the source (Figure \ref{fig:Time_resolved_spectral_study}).
\\[6pt]
Temporal analysis was carried out on the 0.1 $-$ 100 Hz RMS normalized PDS in the 4.0 $-$ 20.0 keV energy range to probe the temporal behaviour of the source. The RMS normalized PDS was dominated by Poisson noise and showed the presence of broadband noise components, which were fit with broad Lorentzians. No QPO features were detected. 
\\[6pt]
A Type I thermonuclear X-ray burst, as observed in several NS-LMXBs \citep{Chen1997}, was detected in the \textit{LAXPC-20} lightcurve of \textit{Observation 3}. This burst showed a double peak profile in the energy ranges 6.0 $-$ 9.0, 9.0 $-$ 12.0 and 12.0 $-$ 20.0 keV, that is interpreted to be a PRE burst \citep{Lewin1984}. The burst rise time and e-folding time, determined following the method adopted by \citet{Galloway2008}, are found to be 0.6 s and 5.6 s, respectively. GX 3$+$1 is known to exhibit three types of Type I thermonuclear X-ray bursts: normal bursts \citep{Asai1993}, short helium flashes \citep{denHartog2003}, and on one occasion, a superburst \citep{Kuulkers2002}. Based on the burst characteristics (burst rise time of 0.6 s and e-folding time of 5.6 s), we deem the burst studied in this work to be a short type helium flash. The duration of the burst is dictated by the amount of hydrogen mixed with fuel of the flash, which is mostly helium; the short e-folding timescale of this burst being caused due to a runaway thermonuclear process, whose fuel is pure helium. The burst attains a peak unabsorbed flux value of (2.87 $\pm$ 0.13) $\times$ $10^{-8}$ erg cm$^{-2}$ s$^{-1}$ in the 4.0 $-$ 20.0 keV energy range in $\sim$3 s since the start of the burst. This is consistent with the result (3.3 $\pm$ 0.6 $\times$ 10$^{-8}$ erg cm$^{-2}$ s$^{-1}$) obtained by \citet{Kuulkers2002}. However, it is to be noted that the burst studied by \citet{denHartog2003} was a superburst with a much lager decay time of $\sim$1.6 h. Since, during the peak of the burst, the observed burst flux is near or at the Eddington limit, the Eddington flux ($F_{Edd}$) is inferred to be 2.87 $\pm$ 0.13 $\times$ $10^{-8}$ erg cm$^{-2}$ s$^{-1}$. Time resolved spectral analysis of the burst carried out using the classical background approach, which uses a fixed pre-burst spectral model with the addition of a blackbody model ($\mathtt{bbody}$), shows the burst to reach a maximum radius of 20.07 $\substack{+1.21\\-1.08}$ km, caused due to the expansion of photosphere of the neutron star. This then touches down to 9.19 $\substack{+0.97\\-0.82}$ km. The rapid expansion and contraction of the blackbody radius along with anti-correlated blackbody temperature confirms the PRE nature of the burst \citep{Galloway2008}. Assuming the burst to be fueled by pure helium, the Eddington luminosity was calculated to be 2.7 $\substack{+0.02\\-0.09}$ $\times$ 10$^{38}$ erg s$^{-1}$. This is consistent with the Eddington luminosities (3.0 $\times$ 10$^{38}$ erg s$^{-1}$) of LMXBs for such bursts seen in globular cluster sources with known distances. The source distance was estimated to be 8.84 $\substack{+0.04\\-0.21}$ kpc using Eddington luminosity (Eq. 5), assuming isotropic radiation of the burst, which is higher compared to the value (6.1 $\pm$ 0.1 kpc) reported by \citet{Kuulkers2000}. Taking anisotropy (0.5 $<$ $\xi_b$ $<$ 2) into account, \citet{Kuulkers2000} estimated the source distance to be between 3 $-$ 7 kpc. Further, \citet{Galloway2020} calculated the source distance to be 5.8 $\pm$ 0.7 kpc, accounting for anisotorpy and assuming the neutron star radius to be 11.2 km. Considering anisotropy constant, $\xi_b = 0.76$ and assuming the neutron star radius to be 9.19 $\substack{+0.97\\-0.82}$ km, we have estimated the source distance to be 10.17 $\substack{+0.07\\-0.18}$ kpc. Further, with neutron star mass, radius and source distance of 1.4 M$_{\odot}$, 9.19 $\substack{+0.97\\-0.82}$ km and 10.17 $\substack{+0.07\\-0.18}$ kpc, respectively, the Eddington mass accretion rate ($\dot{m}_{Edd}$) and pre-burst mass accretion rate ($\dot{m}$) were calculated to be 5.02 $\times$ $10^{-8}$ $M_{\odot}$ y$^{-1}$ and 4.0 $\times$ 10$^{-9}$ $M_{\odot}$ y$^{-1}$, respectively, from the Eddington flux (2.87 $\times$ $10^{-8}$ erg cm$^{-2}$s$^{-1}$) and the pre-burst flux (0.23 $\times$ $10^{-8}$ erg cm$^{-2}$s$^{-1}$). This corresponds to pre-burst mass accretion rate of 0.08 times the Eddington mass accretion rate ($\dot{m}$ = 0.08 $\dot{m}_{Edd}$), as is expected from atoll source in the lower banana branch which the source exhibits in its HID during \textit{Observation 3}, when the burst occurred. 
\\[6pt]
It is shown that the frequency of burst oscillations detected in Type I thermonuclear X-ray bursts exhibited by NS-LMXBs can be inferred as the spin frequency of the neutron star \citep{Muno2002b, Galloway2008, Watts2012}. With this in view, an  in-depth systematic temporal analysis was carried out to look for burst oscillations in the 4.0 $-$ 20.0 keV energy range. Burst oscillation search in the 1s FFT window showed the presence of local maxima at 317 Hz during burst rise and 338 Hz after the touchdown phase. These detections had a significance of $\sim$2.5$\sigma$ which may indicate that they arise due to noise. Since the detections was not significant compared to detection of burst oscillations in other sources, the upper limit of the fractional RMS amplitude of the oscillations was also calculated during different stages of the burst, following \citet{Kuulkers2000}. Its values at burst start, burst maximum and right after the radius expansion phase, were found to be 7$\%$, 5$\%$ and 6$\%$, respectively. In comparison, \citet{Kuulkers2000}, calculated the upper limit of the fractional RMS amplitude to be 24$\%$, 7$\%$ and 10$\%$, during a PRE burst with data from the \textit{RXTE} mission. However, it is to be noted that this was done in the 8 $-$ 60 keV range using 2s long FFT windows. If the oscillations are indeed present, their low amplitudes suggests that these are surface mode or cooling wake oscillations \cite[see e.g.][]{Watts2012,Ootes2017}
\section{Conclusions}
\label{sec:conclusions}
The NS-LMXB source GX 3$+$1 was studied over four observations using \textit{SXT} and \textit{LAXPC} instruments on-board \textit{AstroSat}. The source was found to trace the banana branch in its HID indicating that it was in the soft spectral state during all the four observations. Broadband spectral analysis in the 0.7 $-$ 20.0 keV energy range revealed that the spectrum of the source could be adequately modelled with an absorbed $\mathtt{blackbody}$ model and a $\mathtt{powerlaw}$. This yielded a blackbody radius of $\sim$7.74 km and a mass accretion rate of 2.19 $\times$ 10$^{-9}$\(M_\odot\) y$^{-1}$, assuming a source distance of 6.1 kpc. A Type I X-ray burst was detected in \textit{Observation 3}. From a 0.1 s lightcurve in the 4.0 $-$ 20.0 keV energy range, the burst rise time and e-folding time were found to be 0.6 s and 5.6 s, respectively. Time resolved spectral analysis of the burst showed that the burst could be adequately modeled using the classical approach which requires a fixed pre-burst model and a black body component. An anti-correlation between the temperature and the radius of black body was observed, showing that source underwent a PRE during the burst. The radius of the emitting blackbody at touchdown was calculated to be 9.19 $\substack{+0.97\\-0.82}$ km. Further, using the peak burst flux, the Eddington luminosity of the source was found to be 2.7 $\substack{+0.02\\-0.09}$ $\times$ 10$^{38}$ erg s$^{-1}$ and the source distance was estimated to be 10.17 $\substack{+0.07\\-0.18}$ kpc. The pre-burst mass accretion rate was calculated to be 4.0 $\times$ 10$^{-9}$ $M_{\odot}$ y$^{-1}$ which corresponds to 0.08$\dot{m}_{Edd}$, as is expected from atoll sources in the lower banana branch. Temporal analysis of the source in all four observations showed the presence of broadband noise. The upper limits of the fractional RMS amplitude at the burst start, burst maximum and right after the radius expansion phase, were calculated to be 7$\%$, 5$\%$ and 6$\%$, respectively.   
\section*{Acknowledgements}
The authors thank the \textit{SXT} and \textit{LAXPC} POC teams at Tata Institute of Fundamental Research (TIFR), Mumbai, India for the timely release of data and providing the necessary software tools. This publication uses data from the \textit{AstroSat} mission of the Indian Space Research Organisation (ISRO), archived at the Indian Space Science Data Centre (ISSDC). This work has made use of software provided by HEASARC. The authors acknowledge the financial support (No. DS-2B-13013(2)/9/2019-Sec.2 dated 2019 April 29) of ISRO under \textit{AstroSat} Archival Data Utilization Program. One of the authors (SBG) thanks the Inter-University Centre for Astronomy and Astrophysics (IUCAA), Pune, India for the Visiting Associateship. We thank the anonymous referee for his/her persistent valuable suggestions/comments that improved content of the manuscript drastically. We also thank Dr. Aru Beri, Department of Physics, Indian Institute of Science Education and Research, Mohali, India and Mr. Pinaki Roy for providing the code to estimate fractional RMS amplitude of X-ray burst oscillations. We, specially thank Mr. Pinaki Roy for the valuable discussions and suggestions regarding the burst analysis and estimation of fractional RMS amplitude of X-ray burst oscillations. 
\section*{Data availability}
The data utilised in this article are available at \textit{AstroSat}$-$ISSDC website (\url{http://astrobrowse.issdc.gov.in/astro_archive/archive/Home.jsp)}. The software used for data analysis is available at HEASARC website (\url{https://heasarc.gsfc.nasa.gov/lheasoft/download.html)}.
\section*{ORCID IDs}
Neal Titus Thomas: https://orcid.org/0000-0001-9460-3264
\newline Shivappa B. Gudennavar: https://orcid.org/0000-0002-9019-9441
\newline S. G. Bubbly: https://orcid.org/0000-0003-1234-0662
\bibliography{bibliography}{}
\bibliographystyle{\mnras}
\label{lastpage}
\end{document}